\documentclass[twocolumn,apjl]{emulateapj}

\usepackage{natbib}
\usepackage[fleqn]{amsmath}
\usepackage[usenames, dvipsnames]{color}
\usepackage{booktabs}

\definecolor{pink}{rgb}{0.858, 0.188, 0.478}
\definecolor{green}{rgb}{0.5, 0.9, 0.6}
\definecolor{cyan}{rgb}{0, 0.858, 0.858}

\shorttitle{Local Black Hole Mass Function}
\shortauthors{Mutlu Pakdil et al.}

\begin{document}
\title{THE LOCAL BLACK HOLE MASS FUNCTION DERIVED FROM THE $M_{BH}-P$ AND THE $M_{BH}-n$ RELATIONS}

\author{Burcin Mutlu Pakdil$^{1,2}$, Marc S.\ Seigar$^{1}$, and Benjamin L.\ Davis$^{3,4}$}
\footnotetext[1]{Department of Physics and Astronomy, University of Minnesota Duluth, Duluth, MN 55812}
\footnotetext[2]{Minnesota Institute for Astrophysics, University of Minnesota, Twin Cities, MN 55455}
\footnotetext[3]{Arkansas Center for Space and Planetary Sciences, University of Arkansas, Fayetteville, AR 72701}
\footnotetext[4]{Department of Physical Sciences, Arkansas Tech University, Russellville, AR 72801}

\begin{abstract}
We present a determination of the supermassive black hole (\texttt{SMBH}) mass function for early- and late-type galaxies in the nearby universe ($z < 0.0057$), established
from a volume-limited sample consisting of a statistically complete collection of the brightest spiral galaxies in the southern hemisphere. The sample is defined by limiting
luminosity (redshift-independent) distance, $D_{L}=25.4$ Mpc, and a limiting absolute B-band magnitude, $\mathfrak{M}_{B}=-19.12$. These limits define a sample of 140 spiral, 30 
elliptical (E), and 38 lenticular (S0) galaxies. We established the S\'{e}rsic index distribution for early-type (E/S0) galaxies in our sample. \citet{Davis2014} established the pitch angle 
distribution for their sample, which is identical to our late-type (spiral) galaxy sample. We then used the pitch angle and the S\'{e}rsic index distributions in order to estimate
the \texttt{SMBH} mass function for our volume-limited sample. The observational simplicity of our approach relies on the empirical relation between the mass of the central
\texttt{SMBH} and the S\'{e}rsic index \citep{Graham2007} for an early-type galaxy or the logarithmic spiral arm pitch angle \citep{Berrier2013} for a spiral galaxy. 
Our \texttt{SMBH} mass function agrees well at the high-mass end with previous values in the literature. At the low-mass end, while inconsistencies exist in previous works that still need
to be resolved, our work is more in line with expectations based on modeling of black hole evolution.
\end{abstract}


\section{Introduction}
Rapid technological developments since the early 1990s have provided us with an enormous amount of
information about the existence of Supermassive Black Holes (\texttt{SMBH}s; $M_{bh}\thicksim 10^{5}-10^{9} M_{\sun}$)
in almost all galaxies \citep{KormendyRichstone1995}. Studies have shown that there is a correlation between the \texttt{SMBH} mass and a number
of measurable features of the host galaxy due to the interaction between the \texttt{SMBH} and its surroundings. Some
of the properties known to correlate well with the \texttt{SMBH} mass are the bulge luminosity  
\citep[$L_{Bulge}$;][]{KormendyRichstone1995,MarconiHunt2003}, the bulge mass \citep[$M_{Bulge}$;][]{KormendyRichstone1995, MarconiHunt2003, HaringRix2004},
the mean velocity dispersion $(\sigma)$ of the bulge stars \citep{Ferrarese2000, Gebhardt2000},
the S\'{e}rsic index $(n)$ of the major-axis surface brightness profile \citep{GrahamDriver2007}, and the pitch angle $(P)$ of spiral arms in disk galaxies 
\citep{Seigar2008, Berrier2013}. 

Over the past decade, the number of galaxies with secure mass estimates has increased, because studies have revealed new scaling relations
and revised the existing ones, thus improving our understanding of galaxy-black hole coevolution. The substructures in the most commonly 
cited black hole scaling relations (e.g. $M_{bh}$-$L_{Bulge}$, $M_{bh}$-$\sigma$, $M_{bh}$-$M_{Bulge}$) are reported due to the barred galaxies and/or pseudobulges. 
The true nature of galaxy evolution in different galaxy types still needs to be resolved.

A common practice with these correlations is to estimate the mass function of the central \texttt{SMBH}s (\texttt{BHMF}) in the local universe 
\citep[e.g.][]{Salucci1999, YuTremaine2002, Marconi2004, Shankar2004, Graham2007, Vika2009, Davis2014}. A robust \texttt{BHMF} helps to describe
the evolution of the \texttt{SMBH} distribution and provides important constraints on the coevolution of the quasar and black hole populations. 
The most well-known theoretical constrains are on the integrated emissivity of the quasar population, integrated mass density of black holes, and 
the average black hole accretion rate \citep{Soltan1982, Fabian1999, Elvis2002, Shankar2009}. A comparison among the recent local \texttt{BHMF} 
estimates derived from different scaling relations can be seen in Figure 5 of \citet{Shankar2009}. Most of these studies use an analytic approach, 
which combines the measurements of the galaxy luminosity or velocity function with one of the \texttt{SMBH} scaling relations as outlined 
by \citet{HaringRix2004}. These studies use some assumptions of the morphological type fractions and the bulge-to-total luminosity ($B/T$) ratios.
The sensitivity of the low-mass end of the \texttt{BHMF} based on these assumptions is well presented in Figure A2 of \citet{Vika2009}.
Recently, \citet{Davis2014} estimated the \texttt{BHMF} by using the \texttt{SMBH} mass versus spiral arm pitch angle relation for a nearly complete sample of local spiral galaxies 
in order to produce reliable data for the low-mass end of the local \texttt{BHMF}. In this paper, we aim to estimate a local \texttt{BHMF} for 
all galaxy types within the same volume limits in order to complement this late-type \texttt{BHMF}. Therefore, we used the identical sample selection 
criteria used by \citet{Davis2014}.

The structure of the paper is as follows: in Section 2, we discuss the robustness of the $M_{bh}$-$P$ relation for late-type
galaxies and the $M_{bh}$-$n$ relation for early-type galaxies (E/S0). In Section 3, we describe our sample selection and its completeness.
In Section 4, we present our methodology for estimating \texttt{BHMF}. We first describe how we measure the S\'{e}rsic indeces and how we establish the S\'{e}rsic index 
distribution for the early-type galaxies in our sample. Then, we show our determination of the local \texttt{BHMF} from
 the S\'{e}rsic index distribution for the early-types and the pitch angle distribution for the late-types. Finally, in Section 5 we compare our results 
to the previous works. 

A cosmological model with $\Omega_{\Lambda}=0.691$, $\Omega_{M}=0.307$, $\omega_{b}=0.022$ and $h_{67.77}=H_{o}/$(67.77 km s$^{-1}$ Mpc$^{-1}$)
is adopted throughout this paper.

\section{$M_{bh}$-$P$ Relation and $M_{bh}$-$n$ Relation}
A common conclusion based on observational data is that \texttt{SMBH}s are associated with the
mass of the central bulge in the host galaxy. The $M_{bh}$-$M_{Bulge}$, $M_{bh}$-$L_{Bulge}$, and $M_{bh}$-$n$ relations all depend
on the success of the measurements of the central bulge. In late type galaxies, there can be difficulties when it comes to
isolating the central bulge from other components of galaxies (e.g. bars, disc, and spiral arms). In the study of disc galaxies,
a standard practice is to assume a fixed value of $B/T$ ratio. This introduces a bias on a \texttt{BHMF} such that \texttt{SMBH} mass is over-estimated 
in the late-type disc galaxies and underestimated in early-type disc galaxies \citep{Graham2007}. Another approach is to use the average $B/T$ ratios 
derived from $R^{1/n}$-bulge $+$ exponential-disc decompositions \citep{Graham2007}, which requires heavy image processing tools. The large scatter
in these relations to estimate \texttt{SMBH} mass can be traced back to the complexity of the decomposition in late-type galaxies, particularly in barred galaxies.
The $M_{bh}$-$\sigma$ relation has had considerable success in estimating \texttt{SMBH} masses in many galaxies.  However, it requires spectroscopic measurements, which are 
observationally expensive and depend on the spectroscopic bandwidth.  Furthermore, a careful approach is needed such that a consistent bulge region is always sampled for the measurement of $\sigma$.
Similar to the above relations, measuring $\sigma$ is more complex for disc galaxies than it is for elliptical galaxies because the velocity dispersion from the motion of 
disc and bar is coupled with $\sigma$ and they need to be handled properly \citep{Hu2008}.

Among other relations, the $M_{bh}$-$P$ relation seems promising for late-type galaxies. \citet{Berrier2013} established a linear $M_{bh}$-$P$ relation 
for local spiral galaxies as $log(M/M_{\sun})=(8.21\pm0.16)-(0.062\pm0.009)|P|$ with a scatter less than 0.48 dex in all of their samples. 
This is lower than the intrinsic scatter ($\approx0.56$ dex) of the $M_{bh}$-$\sigma$ relation,
using only late-types \citep{Gultekin2009}. The $P$ derived \texttt{SMBH} mass estimates also seem to be consistent in galaxies with pseudobulges, where other
relations seem to fail \citep{Berrier2013}. Although there are obvious advantages in using the $M_{bh}$-$P$ relation in late-type galaxies (see Discussion in \citet{Berrier2013}) , 
one needs to use a complimentary relation for elliptical and S0 galaxies since the $M_{bh}$-$P$ relation is just applicable for spiral galaxies.
Figure 6 in \citet{Berrier2013} presents evidence that $n$ and $P$ derived mass estimates are compatible for non-barred galaxies, and 
a combination of these two approaches (i.e. using S\'{e}rsic index for E/S0 galaxies, and pitch angles for spiral galaxies) may produce a very 
accurate \texttt{BHMF} for all galaxy types by using only imaging data.

\citet{Graham2001} presented evidence that the light concentration of the spheroids correlate well with their \texttt{SMBH} mass, showing that more centrally
concentrated spheroids have more massive black holes. Given that the S\'{e}rsic index, $n$, is essentially a measurement of the central light concentration,
\citet{GrahamDriver2007} found a log-quadratic relation between n and $M_{bh}$: 
\begin{equation}
\log(M_{bh}) = (7.98\pm0.09)+(3.70\pm0.46)\log(\frac{n}{3})-(3.10\pm0.84)[\log(\frac{n}{3})]^2
\end{equation}
with an intrinsic scatter of $\epsilon_{intrinsic}=0.18^{+0.07}_{-0.06}$ dex.

Recently, \citet{Sani2011}, \citet{Vika2012} and \citet{Beifiori2012} failed to recover a strong $M_{bh}$-$n$ relation. \citet{Savorgnan2013} re-investigated and recovered 
the relation using a large collection of literature S\'ersic index measurements using R-band \citep{GrahamDriver2007}, I-band \citep{Beifiori2012}, K-band \citep{Vika2012}, and 3.6\micron
\citep{Sani2011} imaging data. \citet{Savorgnan2013} discussed the systematic effects associated with measuring S\'ersic index in different optical and infrared wavebands.  They concluded that 
the differences expected from measuring S\'ersic index in different wavebands are smaller than the differences expected due to other systematic biases such as one-dimensional decomposition 
versus two-dimensional decomposition, or the differences between measuring the S\'ersic index along a minor axis versus measuring it along a major axis.  Indeed, one migh expect that a S\'ersic index 
measured using a one-dimensional fit (as performed in this paper) to be $\sim$10\% smaller than that measured using a two-dimensional fit \citep{Ferrari2004}.  Furthermore, when measuring S\'ersic 
index in multiple wavebands for the same galaxies, \citet{Savorgnan2013} found that wavelength bias was completely dominated by these other biases, which could be as large as 50\%.  Given, the 
result of \citet{Kelvin2012}, we would expect the S\'ersic index measured at 3.6\micron to be less than 10\% higher than that measured in the $R$-band, which is signifantly smaller than the 50\% 
number given by \citet{Savorgnan2013}.  \citet{Savorgnan2013} excluded the outlying S\'ersic indices, averaged the remaining values, and recovered the $M_{bh}$-$n$ relation by showing that elliptical 
and disc galaxies follow two different linear $M_{bh}$-$n$ relations. They discussed how this relation is consistent with what would be derived by combining the $M_{bh}$-$L_{Bulge}$ and $L_{Bulge}$-$n$ 
relations and how this explains the log quadratic nature of the $M_{bh}$-$n$ relation reported by \citet{GrahamDriver2007}. 

In this paper, we define early-type galaxies as elliptical and S0 galaxies. The sample used by \citet{GrahamDriver2007} was dominated ($\sim89\%$) by elliptical and S0 galaxies. However,
\citet{Savorgnan2013} studied S0 galaxies together with spiral galaxies. Therefore, we used the log quadratic $M_{bh}$-$n$ relation reported by \citet{GrahamDriver2007} to estimate \texttt{SMBH} masses 
in our early-type sample.

\section{Data and Sample Selection} 
\citet{Davis2014} based their selection criterion on the Carnegie-Irvine Galaxy Survey (\texttt{CGS}) \citep{Ho2011}; it is an almost complete sample of 605 nearby 
galaxies in the southern hemisphere. Using the spiral galaxies in this parent sample plus Milky Way, they defined a volume-limited sample which consists of spiral galaxies
within a luminosity (redshift-independent) distance of $25.4$ Mpc and a limiting absolute B-band magnitude of 
$\mathfrak{M}_{B} = -19.12$. We followed the same selection criterion, except also including elliptical and S0 galaxies.  As a result, 
our volume-limited sample consists of 208 host galaxies (30 ellipticals and 38 S0s and 140 spiral galaxies) within a comoving volume 
of $V_{c}=3.37\times10^{4}$ h$^{-3}_{67.77}$ Mpc$^3$ over a lookback time, $t_{L}\leq82.1$ Myr. We then downloaded images of selected galaxies from the \texttt{NASA/IPAC} Extragalactic 
Database (\texttt{NED}).

A complete sample selection is necessary to estimate a meaningful \texttt{BHMF}. Therefore, we checked the completeness of our sample within 
the limits of luminosity distance and absolute B-band magnitude in several ways. First, we compared our sample size with the maximum number of galaxies within these limits. 
Figure 1 shows that the maximum number of galaxies, which is 217, appears at $D_{L}=28.05$ Mpc and $\mathfrak{M}_{B}=-19.37$, whereas our sample consists of 
208 galaxies. While these two limits just differ by $4\%$, using the limiting $\mathfrak{M}_{B} =-19.12$ allows us to include galaxies with dimmer intrinsic
brightness, and helps us to be more complete. 

In addition, we determined the luminosity function in order to check if our volume-limited sample is a fair representation of the local galaxy population 
over the absolute magnitude range $-19.12 \lesssim \mathfrak{M}_{B} \lesssim -23$. The luminosity function is determined as $\phi(\mathfrak{M}_{B}) =\partial N/\partial \mathfrak{M}_{B}$, 
where N is the number of galaxies in our sample in terms of the absolute B-band magnitude and dividing it by the comoving volume of the volume-limited sample. 
This is illustrated in Figure 2, which shows the comparison with the luminosity functions for the overall \texttt{CGS} sample \citep{Ho2011} and the much larger sample of 
\citet{Blanton2003} and \citet{Bernardi2013}. 
The luminosity functions of \citet{Blanton2003} and \citet{Bernardi2013} have been shifted by $B-r=0.67$ mag, the average color of an Sbc spiral \citep{Fukugita1995}, which is roughly 
the median Hubble type of both \texttt{CGS} and our volume-limited sample, and also transformed to $H_{0}=67.77$ km s$^{-1}$ Mpc$^{-1}$. While \citet{Blanton2003} derived the luminosity function 
of $z\approx0.1$ galaxies from Sloan Digital Sky Survey (\texttt{SDSS}) by using the S\'{e}rsic parameters from a 1-D radial surface brightness profile, \citet{Bernardi2013} derived it 
by using the 2-D fits to the whole galaxy image. The overall \texttt{CGS} sample has a luminosity function that agrees quite well with that of \citet{Blanton2003} \citep{Ho2011}. 
However, our galaxy sample has a luminosity function that implies that it was observed in an overdense volume (see red data points in Figure 2). Therefore, we renormalized 
our luminosity function by adding $-0.25$ in the y-axis in order to be consistent with that of \citet{Blanton2003} and \texttt{CGS} (see pink data points in Figure 2). For our \texttt{BHMF} estimation,
we used the same normatization factor (see Section 4.3). In addition, due to the sample selection criterion, our luminosity function does not extend below the magnitude limit of 
$\mathfrak{M}_{B}=-19.12$. This fact is obviously of interest to our \texttt{BHMF} estimation that will be discussed more in Section 5.

Furthermore, we compared the distribution of morphological types of \texttt{CGS} and our sample. Our morphological fractions, $f_{type}$, are as such: $f_{E}=0.14$, $f_{S0}=0.18$, and 
$f_{Spiral}=0.67$. This is in good agreement with the ones ($f_{E}=0.11\pm0.03$, $f_{S0}=0.21\pm0.05$, $f_{Sab+Sbc+Scd}=0.62\pm0.14$) reported by \citet{Fukugita1998}. Moreover,
Figure 3 shows that our volume-limited sample preserves the distribution of morphological types in \texttt{CGS}. In addition, we checked the T-type 
distributions of \texttt{CGS} and our sample. The T-type values are taken from http://cgs.obs.carnegiescience.edu/CGS/database\_tables. The differences between the densities 
of each T-type are always less than 5\% (see Figure 4). 

We used imaging data taken from the \texttt{NASA/IPAC} Extragalactic Database (\texttt{NED}) (see Table 3). The absolute magnitudes were calculated from 
apparent magnitudes, from \textit{HyperLeda} \citep{Paturel2003}, luminosity distances compiled from the mean redshift-independent distance from the \texttt{NED},
and extinction factors in the B-band from \citet{SchlaflyFinkbeiner2011}, as compiled by the \texttt{NED}. We used several different band images for 
our measurements. 

\section{Methodology} 
\subsection{S\'{e}rsic Index Measurement}
In order to have a reliable S\'{e}rsic index measurement for early-type galaxies in our sample, we carefully masked the foreground stars and background galaxies
by using the \texttt{SEXTRACTOR} \citep{BertinArnouts1996}, and determined the centers of the galaxies by using the \texttt{IRAF} task \texttt{IMCNTR}. 
The sky-background flux and its uncertainty were estimated from the mean and standard deviation of five median fluxes that were obtained from small boxes near 
the galaxy free corners of each images, respectively. Then, the surface brightness profiles were extracted using the \texttt{IRAF} task \texttt{ELLIPSE} 
\citep{Tody1986, Jedr1987} with a fixed center and allowing the isophotal position angle and ellipticity to vary. The best S\'{e}rsic bulge $+$ exponential disc
model for S0 galaxies, and the best S\'{e}rsic bulge model for elliptical galaxies were fitted by minimizing $\chi^{2}$ with an iterative procedure. 
The models were derived three times for each galaxy in order to estimate the S\'{e}rsic index error. The uncertainity in the sky-background level was
respectively added and subtracted from the surface brightness profile data in the second and third derivation (see Figure 5). 
This method for estimating the errors on the model parameters was also used by \citet{DeJong1996}. When fitting the profiles, seeing effects are particularly
relevant when the ratio between the \texttt{FWHM} of the seeing and the effective half-light radii $R_{e}$ of the S\'{e}rsic model is small \citep{Grahamdisk2001}.
When $R_{e}/$\texttt{FWHM} $> 2$, the difference between the measured S\'{e}rsic index and the actual S\'{e}rsic index is typically small, as explained 
by \citet{Grahamdisk2001}. For our sample, all the derived bulge values for $R_{e}$ are greater than 1\arcsec, and the ratio $R_{e}/$\texttt{FWHM} is 
greater than 2 (see Table 3, Column 6). The results of the best-fitting S\'{e}rsic bulge model for elliptical galaxies and the best-fitting S\'{e}rsic bulge $+$ exponential disc 
model for S0 galaxies are shown in Figure 6 and 7, respectively.

We successfully completed the S\'{e}rsic index measurements for all 68 galaxies in our sample.
Before proceeding, we note that Equation 1 was constructed in the R-band \citep{GrahamDriver2007}, while our data ranges from the R-band to 4.6$\micron$. The structural parameters of 
a galaxy may vary with wavelength due to the radial variations in stellar population and/or dust obscuration \citep{Kelvin2012}. This may result in different values for S\'ersic index in 
different wavelengths. However, the local early-type galaxies mostly have fairly small color gradients \citep[e.g.][]{Peletier1990,Taylor2005}. Using similar fitting method to ours (S\'{e}rsic 
bulge model for ellipticals; S\'{e}rsic bulge $+$ exponential disc model for disc galaxies), \citet{McDonald2011} found that the S\'ersic indices of elliptical and S0 galaxies show no 
significant variation across optical and NIR wavelengths. In order to quantify how photometric and structural parameters of a galaxy vary with wavelength, recent studies used 2D single 
S\'ersic fits and reported that galaxies with different S\'ersic indices and colors follow different trends with wavelength \citep[e.g.][]{Kelvin2012, Vulcani2014, Kennedy2015}. Their 
common result is that high-n galaxies remain relatively stable at all wavelengths. These high-n galaxies roughly correspond to our early-type sample. However, it is worth mentioning that 
the measurement of the S\'ersic index in these recent studies are different to ours: they used a single S\'ersic profile fit for all galaxies and made no attempt to remove objects 
for which a two-component fit would be more appropriate. Therefore, single S\'ersic index wavelength dependence mostly gives information about bulge and disc properties of a galaxy \citep{Kennedy2016}. 
For example, \citet{Vulcani2014} attributed the lack of variation in S\'ersic index with wavelength for red galaxies to the fact that they principally comprise one-component objects (i.e. ellipticals) 
or two-component galaxies in which the components possess very similar colors, i.e. S0s. Although we can get some insight for the (disc-less) elliptical galaxies, the single S\'ersic galaxy 
model is not suitable for quantifying possible changes with wavelength to S\'ersic indices of bulges in S0 galaxies. Therefore, following the work of \citet{McDonald2011}, we did not apply any 
corrections to our S\'{e}rsic index measurements. All measured data for individual early-type galaxies in our sample are listed in Table 3.
 
\subsection{S\'{e}rsic index distribution}
As a result of S\'{e}rsic index measurements, we had three S\'{e}rsic index estimates ($n_{i}$) for each of our 68 galaxies. We used two independent ways in order to find the best fit
probability density function (\texttt{PDF}) to our data. 

First, we employed a nominal \textit{binless} histogram, which is identical to the method in \citet{Davis2014}, in order to create the S\'{e}rsic index distribution.
We modeled each data point twice as a normalized Gaussian, where the mean is the average S\'{e}rsic index values $<n_{i}>$ and the standard deviation is the standard deviation 
of $n_{i}$ , $\sigma_{<n_{i}>}$. The S\'{e}rsic index distribution is obtained by a normalized sum of Gaussian values. Then, we repeated the same modeling, but this time 
the mean is the average logarithmic value of $n_{i}$, $<\log n_{i}>$, and the standard deviation is the standard deviation of $\log n_{i}$, $\sigma_{<\log n_{i}>}$. From the 
resulting S\'{e}rsic index distributions, we were able to compute the statistical standardized moments of a probability distribution; mean ($\mu$), standard deviation (stdev), 
skewness, and kurtosis. The two distributions give us almost the same statistical standardized moments: $\mu=3.10 (3.10)$, $stdev=1.38 (1.39)$, $skewness=0.95 (0.95)$, and 
$kurtosis=4.17 (4.18)$, where the numbers in parentheses refer to the distribution derived from $<n_{i}>$ and $\sigma_{<n_{i}>}$. We used the \texttt{MATLAB} code \texttt{PEARSPDF} 
to perform our \texttt{PDF} fitting. To explore the uncertainty in our \texttt{PDF} fit, we used a bootstrapping process. The random number generator \texttt{NORMRND} in \texttt{MATLAB} 
was used for sampling (with replacement) from the original 68 data points, using the mean as $<\log n_{i}>$ and the standard deviation as $\sigma_{<\log n_{i}>}$. 
The statistical standardized moments for one thousand data sets containing \text{68} data points each were individually calculated. 
This gave one thousand new estimates for each of the parameters ($\mu$, stdev, skewness, and kurtosis). Then, the median and the standard deviation of these new estimates gave us 
the uncertainty on the \texttt{PDF} fitting: $\mu=3.12\pm0.02$, $stdev=1.40\pm0.04$, $skewness=0.92\pm0.03$, $kurtosis=3.87\pm0.30$. 

Then, we used the \texttt{MATLAB} code \texttt{ALLFITDIST}, which fits all valid parametric probability distributions to the data and returns the fitted distributions based on 
the Bayesian information criterion. As a result, the gamma distribution function is given as a best \texttt{PDF} fit, with $\mu=3.11$, $variance=1.84$, shape $a=5.26\pm0.51$, 
scale $b=0.59\pm0.06$. The resulting S\'{e}rsic distribution and its \texttt{PDF} fits are illustrated in Figure 8.

\subsection{Estimating BHMF}
The local \texttt{BHMF} is formulated as
\begin{equation}
\phi(\log(M_{bh}))=\frac{\partial N}{\partial \log(M_{bh})}=\frac{\partial N}{\partial x}\frac{\partial x}{\partial \log(M_{bh})}=\phi(x)\frac{\partial x}{\partial \log(M_{bh})}
\end{equation}
where $N$ is the number of galaxies, $x$ is pitch angle $P$ for late-type galaxies and S\'{e}rsic index $n$ for early-type galaxies, and $M_{bh}$ is \texttt{SMBH} mass.
For the early-type galaxies, the S\'{e}rsic index measurements for the volume-limited sample give us the S\'{e}rsic index function
$\phi(n)=\frac{\partial N}{\partial n}$; and $\frac{\partial n}{\partial \log(M_{bh})}$ can be evaluated by taking the derivative of Equation 1 as follows:
\begin{equation}
\frac{d\log(M_{bh})}{dn}=\frac{(3.70\pm0.46)}{n\ln(10)}-\frac{2(3.10\pm0.84)\log(\frac{n}{3})}{n\ln(10)}
\end{equation}
As a result, we get the following equation:
\begin{equation}
\phi(\log(M))=\phi(n)[\frac{(3.70\pm0.46)}{n\ln(10)}-\frac{2(3.10\pm0.84)\log(\frac{n}{3})}{n\ln(10)}]^{-1}
\end{equation}
Using Equation 5 and dividing by a local comoving volume of $V_{c}=3.37\times10^{4}$ $h^{-3}_{67.77}$ $Mpc^{3}$, the S\'{e}rsic index distribution was converted into the \texttt{BHMF} 
for the early-type galaxies.

In order to estimate the error in the \texttt{BHMF}, we ran a Markov Chain Monte Carlo (\texttt{MCMC}) sampling of the \texttt{BHMF}. The sampling uses $10^{5}$ realizations of 
the S\'{e}rsic index distribution based on the errors in the previous section. The S\'{e}rsic index distributions were randomly generated from the parameters that define 
the \texttt{PDF}, assuming that they are normally distributed with the $1\sigma$ uncertainties given by the estimated errors. The uncertainties in the $M_{bh}$-$n$ relation are also 
allowed to vary as a Gaussian distribution around the fiducial values. We first estimated the \texttt{BHMF} without assuming any errors, then we allowed the listed errors 
(four parameters in the \texttt{PDF} fit $+$ three parameters in the $M_{bh}$-$n$ relation) to be perturbed individually and collectively. This is illustrated in Figure 9 (left), 
which shows that the S\'{e}rsic index distribution has no impact on the \texttt{BHMF} for $M_{bh}>10^{9}M_{\sun}$ since the mass of the \texttt{SMBH} is fixed for $n > 11.9$. 
The sharp decrease at the high-mass end is the result of the curved nature of the $M_{bh}$-$n$ relation, that predicts a maximum mass which \texttt{SMBH}s have formed \citep{Graham2007}.
The uncertainties in the $M_{bh}$-$n$ relation dominate at this region, softening the high-mass decrease of the \texttt{BHMF}, and thus increasing the total density of the \texttt{BHMF} 
for high masses. 

The error region in the \texttt{BHMF} is estimated by the $16^{th}$ and $84^{th}$ percentile of the $10^{5}$ \texttt{MCMC} realizations, similar to the
method used by \citet{Marconi2004}, where the $16^{th}$ and $84^{th}$ percentiles indicate the $1\sigma$ uncertainties on the logarithm of the local \texttt{BHMF}. 
In order to deal with the intrinsic scatter in the $M_{bh}$-$P$ relation, \citet{Davis2014} used the method described in Equation 3 in the paper of \citet{Marconi2004}. However, we did 
not adopt this method for our early-type \texttt{BHMF}. \citet{Graham2007} discussed that the intrinsic scatter in the $M_{bh}$-$n$ relation is not Gaussian; and the removal of the two highest
mass \texttt{SMBH}s converts the $M_{bh}$-$n$ relation into one with zero intrinsic scatter. In estimating the \texttt{BHMF} derived from the $M_{bh}$-$n$ relation, \citet{Graham2007}
did not apply any correction for the intrinsic scatter, and neither did we. Finally, we obtained our best estimate of the early-type \texttt{BHMF} by merging all 
the random realizations of the \texttt{BHMF}s and considering the $16th$, $50th$, and $84th$ percentile levels (see the right panel in Figure 9).
We note that the early-type \texttt{BHMF} is normalized by adding $-0.25$ in the y-axis, which corrects for the overdensity in our selected volume.

In order to estimate the local \texttt{BHMF} for all galaxy types, following Equation 3, we also run the \texttt{MCMC} realizations of the \texttt{BHMF} for the spiral galaxies, but 
this time using the pitch angle distribution that was derived by \citet{Davis2014}. Note that \citet{Davis2014} estimated possible \texttt{SMBH} masses from the $M_{bh}$-$P$ relation by using 
the \texttt{MCMC} sampling and then fitted a \texttt{PDF} model to derive the late-type \texttt{BHMF}. In this paper, we used the best-fit \texttt{PDF} model for the pitch angle distribution 
derived by \citet{Davis2014}, and then used Equation 3 by adopting the method used by \citet{Marconi2004} to estimate the late-type \texttt{BHMF} by considering the $16th$, $50th$, and $84th$ 
percentile levels of the \texttt{MCMC} realizations. Similar to our early-type MCMC sampling, we assumed that the input parameters ($\mu$, stdev, skewness, kurtoisis) of the \texttt{PDF} fit 
and the uncertainities in the $M_{bh}$-$P$ relation are Gaussian distributed around the fiducial values. Then, we merged all random realizations of \texttt{BHMF}s from the early-type and 
spiral galaxies. Figure 10 shows our best estimate of the local \texttt{BHMF} obtained by merging all random realizations and considering the $16th$, $50th$, and $84th$ percentile levels. 
The late-type \texttt{BHMF} and the early-type \texttt{BHMF} are also shown in Figure 10 to help visualize how the early- and late-type samples are being spliced. We note that the our 
\texttt{BHMF} estimates are all normalized by adding $-0.25$ in the y-axis to be able correct for the overdensity in our survey volume. The plotted data for Figure 9 (right) and Figure 10 are listed 
for convenience in Table 1.

\subsection{\texttt{SMBH} mass density}
Integrating over the mass functions, we derived the local mass density of \texttt{SMBH}s which gives $1.74^{+0.79}_{-0.60}\times10^{5}$ h$^3_{67.77}$ M$_{\sun}$ Mpc$^{-3}$ for early-type and 
$2.04^{+1.16}_{-0.75}\times10^{5}$ h$^3_{67.77}$ M$_{\sun}$ Mpc$^{-3}$ for all-type galaxies. For reference, \citet{Graham2007} and \citet{Vika2009} reported 
$3.99\pm1.54\times10^{5}$ h$^{3}_{67.77}$ M$_{\sun}$ Mpc$^{-3}$ and $7.25\pm1.18\times10^5$ h$^3_{67.77}$ M$_{\sun}$ Mpc$^{-3}$ for the \texttt{SMBH} mass density in the local all-type galaxies, respectively.
In terms of the critical density of the universe, we obtained $\Omega_{BH,total}=1.61^{+0.91}_{-0.59}\times10^{-6}$ h$_{67.77}$.
This implies that $0.007^{+0.005}_{-0.003}$ h$^{3}_{67.77}$ percent of the baryons are contained in \texttt{SMBH}s at the centers of galaxies in the local universe (see Table 2).

\section{Discussion}
Figure 11 shows the comparison of our early-type \texttt{BHMF} with previously estimated early-type \texttt{BHMF}s \citep{Graham2007, Marconi2004, Vika2009}.
Our early-type \texttt{BHMF} is expected to be consistent with that of \citet{Graham2007} within the uncertainties, since they are both derived from the same $M_{bh}$-$n$ relation.
The data points are in overall good agreement within their uncertainties. There is an apparent disagreement below $M_{bh} < 10^{6.5}M_{\sun}$, which corresponds to $n \approx 1.5$ and the region between
$10^{8}M_{\sun} < M_{bh} < 10^{8.75}M_{\sun}$. \citet{Graham2007} defined early-type galaxies as $\frac{B}{T}>0.4$ and used the GIM2D$-$derived $n$ values \citep{Allen2006}, which were obtained 
from the logical filter for S\'{e}rsic $+$ exponential catalog. For galaxies with $n < 1.5$, this logical filter classifies galaxies as pure disk and therefore fits them with a single component.  
However, we obtained $1 < n < 1.5$ for seven S0 galaxies but still performed a two-component fit. As a result, our \texttt{BHMF} has higher density for the low mass end ($M_{bh} < 10^{6.5}M_{\sun}$) 
and lower density for intermediate masses ($10^{8}M_{\sun} < M_{bh} < 10^{8.75}M_{\sun}$). Differences in the definition of early-type galaxies and the profile fitting methodology may explain the 
disagreement between the two \texttt{BHMF}s derived from the same relation. It should also be noted that they used a sample of 1356 early-type galaxies from the Millennium Galaxy Catalogue 
(\texttt{MGC}) in the redshift range of $0.013 < z < 0.18$, and they estimated the \texttt{BHMF} by summing the \texttt{SMBH} mass distribution times an associated space-density weights, 
i.e., $\phi(M)=\sum W(L)M$, where $W(L)=\phi(L)/N(L)$ is constructed for black holes derived from early-type galaxies (defined as $\frac{B}{T}>0.4$). Although the volume of their sample is 
considerably higher than ours, and their sample selection and \texttt{BHMF} estimation method are different from ours, overall their \texttt{BHMF} is consistent with our findings.

We also compared our \texttt{BHMF} with the work of \citet{Vika2009}. They used the sample identical to that of \citet{Graham2007}, except they included the galaxies with $\mathfrak{M}_{B}>-18$, 
indicating the data from this region is unreliable. They used the linear $M_{bh}$-$L_{Bulge}$ relation reported by \citet{Graham2007Lum} with dust correction to their sample. Other than using 
the $M_{bh}$-$L_{Bulge}$ relation to derive the \texttt{BHMF}, their \texttt{BHMF} estimation method is identical to that of \citet{Graham2007}. However, their \texttt{BHMF} does not agree 
well with that of \citet{Graham2007}, or with ours. They discussed the probable reasons for the discrepancy between theirs and that of \citet{Graham2007} (see Section 3.1 in 
\citet{Vika2009}). In addition, \citet{Graham2013} recently revised the $M_{bh}$-$L_{Bulge}$ relation and found a log quadratic nature in the $M_{bh}$-$L_{Bulge}$ relation, which is also 
expected from the linear nature of the two distinct $L_{Bulge}$-$n$ relations for elliptical galaxies and bulges, and the curved $M_{bh}$-$n$ relation. This may explain the discrepancy 
between the \texttt{BHMF} derived from the linear $M_{bh}$-$L_{Bulge}$ relation and the one derived from the curved $M_{bh}$-$n$ relation. 

In addition, we compared our \texttt{BHMF} with that of \citet{Marconi2004}. They estimated the local \texttt{BHMF} for early-type galaxies based on \texttt{SDSS} sample of \citet{Bernardi2003},
by using the linear $M_{bh}$-$L_{Bulge}$ and $M_{bh}$-$\sigma$ relations reported by \citet{MarconiHunt2003} assuming the same intrinsic dispersion. They also derived the local 
\texttt{BHMF} for early-type galaxies obtained from different galaxy luminosity functions, in different photometric bands. All their local \texttt{BHMF}s for early-type 
galaxies are in remarkable agreement with ours within the uncertainities. However, they reported a discrepancy at $M_{bh} < 10^{8} M_{\sun}$ between the \texttt{BHMF} derived with the 
\citet{Bernardi2003} luminosity function and the others (see Figure 1b in \citet{Marconi2004}). They considered this discrepancy as insignificant because this is the region where authors 
adopted different functional forms to fit the data to extrapolate luminosity functions of early-type galaxies. Our early-type \texttt{BHMF} agrees more with the one derived from the sample of
\citet{Bernardi2003} at $M_{bh} < 10^{8} M_{\sun}$ than the others.

Figure 12 shows the comparison between our \texttt{BHMF} for all galaxy types with those of \citet{Graham2007}, \citet{Vika2009}, and \citet{Marconi2004}.
Overall our \texttt{BHMF} agrees better with that of \citet{Marconi2004} within the uncertainties. It is clear that there is a disagreement between ours and 
those of \citet{Graham2007} and \citet{Vika2009} at the low-mass end. Late-type galaxies have the biggest contribution on the \texttt{BHMF} at the low-mass end (see Figure 10), 
where the S\'{e}rsic index is more difficult to measure due to the complex nature of these late-type galaxies as we explained earlier in this paper. 
It is also worth mentioning that \citet{Vika2009} argued that their \texttt{BHMF} data below 
$\log(M_{bh}/M_{\sun}) = 7.67$ (light blue circles in Figure 12) is not reliable because it is derived from galaxies with $\mathfrak{M}_{B}>-18$. Our entire sample consists of galaxies 
with $\mathfrak{M}_{B}\leq-19.12$. Moreover, \citet{Davis2014} stated a possible bias for the sample of \citet{Vika2009}, pointing the small number of late-type galaxies in their considerably 
larger sample volume (see Section 7 in \citet{Davis2014}). Although our sample does not contain very faint galaxies $(\mathfrak{M}_{B} > -19.12)$, our \texttt{BHMF} results in a higher 
number density for the low-mass end when compared to those of \citet{Vika2009} and \citet{Graham2007}. In addition, other relations ($M_{bh}$-$n$, $M_{bh}$-$L_{Bulge}$, and $M_{bh}$-$\sigma$ relations) 
are not as accurate as the $M_{bh}$-$P$ relation in this mass regime \citep{Berrier2013}.

Finally, Figure 13 shows the comparison between our all-type \texttt{BHMF} with more recent \texttt{BHMF} estimates \citep{Shankar2013b, Sijacki2015}. 
At the high-mass end, it looks as if our \texttt{BHMF} lies between those of \citet{Marconi2004} and \citet{Shankar2013b}, except for the lower mass \texttt{SMBH}s with $M_{bh} < 10^{7}M_{\sun}$.
\citet{Shankar2013b} derived the local \texttt{BHMF} based on the assumption that all local galaxies follow the early-type $M_{bh}$-$\sigma$ relation reported by \citet{McConnellMa2013}. 
As shown in Figure 10, early-type galaxies dominate at the high-mass end, therefore a \texttt{BHMF} derived from a relation for early-type galaxies is expected to be more reliable at the high-mass 
end. Observational uncertainities increase for low mass (late-type) galaxies because measuring $\sigma$ in disc galaxies is not a trivial task and one needs to properly count the contribution 
from the motion of disc and bar that is coupled with the bulge. In addition, the majority of low mass galaxies may host pseudobulges \citep{FisherDrory2011}, and a number of independent groups 
claimed that the properties measured for galaxies with pseudobulges do not follow the typical scaling relations (e.g. $M_{bh}$-$\sigma$, $M_{bh}$-$M_{Bulge}$, $M_{bh}$-$L_{Bulge}$), 
with \texttt{SMBH} masses being often significantly smaller than what is expected by these relations \citep[e.g.][]{Hu2009, Greene2010, Kormendy2011, Beifiori2012}. Therefore, the \texttt{BHMF} 
of \citet{Shankar2013b} (and most of previous ones) likely represents an upper limit on the true local \texttt{BHMF} \citep{Shankar2013b}. To adress this issue, \citet{Shankar2013b} re-estimated 
the \texttt{BHMF} with the same relation, but this time the authors made the odd assumption that Sa galaxies do not host any \texttt{SMBH}s. This assumption likely makes this modified \texttt{BHMF} 
a lower limit on the local \texttt{BHMF} \citep{Sijacki2015}. Our \texttt{BHMF} indeed stays between the \texttt{BHMF} of \citet{Shankar2013b} and the modified one. In the comparison with 
the \texttt{BHMF}s derived from accretion models, the continuity equation models of \citet{Shankar2013a} predict a local \texttt{BHMF} similar to that of \citet{Shankar2013b} when a constant 
Eddington ratio is assumed (see Figure 2 of \citet{Shankar2013b}), and they predict a local \texttt{BHMF} very similar to ours for the highest mass regime when a Eddington ratio is assumed to 
be decreasing as a function of cosmological time (see dot-dashed line in Figure 13).  Finally, when compared with the Illustris Simulation, which is a large scale cosmological simulation with 
the resolution of a (106.5 Mpc)$^3$ volume, our result agrees quite well with their \texttt{BHMF}. At higher masses, the simulation estimate is in a remarkable agreement with our result. 
Similar to the others, disagreements exist at lower masses, and \citet{Sijacki2015} already argued that the simulation results are least reliable at the low-mass end (see Section 3.3 in 
\citet{Sijacki2015}).  In summary, for the intermediate and high mass \texttt{SMBH}s ($M_{bh} > 10^{7} M_{\sun}$), the agreements between our \texttt{BHMF} and those of previous \texttt{BHMF} 
estimates are encouraging. At the low-mass end, inconsistencies exist in the previous work that still need to be resolved, but our work is more in line with the expectations based on 
accretion models \citep{Shankar2013a}, 
favouring steadily decreasing Eddington ratios, and semi-analytic models \citep[e.g.][]{Marulli2008}, which suggest a relatively flat distribution for $M_{bh} \lesssim 10^{8}M_{\sun}$. 
Also, our results at the low-mass end of the \texttt{BHMF} are probably consistent with the claims that the majority of low-mass galaxies contain pseudobulges rather than classical 
bulges \citep{FisherDrory2011}.  This, in turn, may explain why the $M_{bh}$-$P$ produces a tighter relation than the $M_{bh}$-$\sigma$ relation for disc galaxies \citep{Berrier2013}, and therefore 
why our \texttt{BHMF} result shows more promise when compared to expectations from semi-analytical models.  This highlights an important need for properly accounting for the affects of pseudobulges 
in disc galaxies when determining the local \texttt{BHMF}.

\section{Conclusion}
The observational simplicity of our approach and the use of the statistically tightest correlations with \texttt{SMBH} mass, which are the S\'{e}rsic index
for E/S0 galaxies and pitch angle for spiral galaxies, make it straightforward to estimate a local \texttt{BHMF} through imaging data only within 
a limiting luminosity (redshift-independent) distance $D_{L}=25.4$ Mpc $(z=0.00572)$ and a limiting absolute B-band magnitude of $\mathfrak{M}_{B}=-19.12$. 
The inconsistencies at the low-mass end of the local \texttt{BHMF} exist in the previous works that still need to be resolved. We presented our \texttt{BHMF} 
as of a particular interest because it is a nearly complete sample within set limits and provides reliable data, especially for the low-mass end of the local \texttt{BHMF}.

\section*{Acknowledgements}
This research has made use of the \texttt{NASA}/\texttt{IPAC} Extragalactic Database (\texttt{NED}) which is operated by the Jet Propulsion Laboratory, California Institute of
Technology, under contract with National Aeronautics and Space Administration. The authors wish to thank Joel C. Berrier for useful discussion.
MSS wishes to thank the generous support of the University of Minnesota Duluth and the Fund for Astrophysical Research. We also wish to thank the 
anonymous referee whose comments greatly improved the content of this paper.

\clearpage
\begin{figure*}
\begin{center}
\end{center}
\caption{\textbf{Left:} Luminosity distance vs absolute B-band magnitude for all type galaxies (606) found using the magnitude-limiting 
selection criteria ($B_{T}\leq12.9$ and $\delta < 0$). The upper limit absolute magnitude is modeled as the same
exponential in \citet{Davis2014} and is plotted here as the solid black solid line. The dashed red rectangle shows the galaxies in the volume-limited sample.
\textbf{Right:} Histograms showing the number of galaxies (spiral galaxies: blue; early-types (E/S0): red; all-types: pink) contained in the limits of luminosity distance and absolute 
B-band magnitude as the limits are allowed to change on the exponential line based on
the limiting luminosity distance. Note that the peak for all galaxy types appears in the histogram with 217 galaxies at $D_{L}=28.05$ Mpc. Black dashed line represents $D_{L}=25.4$ Mpc 
that is the limit used in \citet{Davis2014}, and gives 208 galaxies (68 E/S0 $+$ 140 spiral), which is very close to the actual peak. Using $D_{L}=25.4$ Mpc allows us to be complete for
dimmer galaxies. Complete volume-limited samples were computed for limiting luminosity distances in increments of 0.001 Mpc.\label{Figure 1}}
\end{figure*}

\begin{figure*}
\caption{Our luminosity function (\texttt{LF}) (red triangles) is shown, in comparison with the \texttt{LF}s for the much larger sample of \citet{Blanton2003}(blue dots),
\citet{Bernardi2013}(green dashed line), and \texttt{CGS}(black stars) \citep{Ho2011}. Our \texttt{LF} implies an overdense region for our volume-limited sample,
therefore we renormalized our \texttt{LF} by adding $-0.25$ in the y-axis. The normalized \texttt{LF} is depicted by pink circles. Note that the \texttt{LF}s of \citet{Blanton2003} and 
\citet{Bernardi2013} have been shifted by $B-r=0.67$ mag, the average color of an Sbc spiral \citep{Fukugita1995}, which is roughly the median Hubble type of both \texttt{CGS} 
and our volume-limited sample, and also transformed to $H_{0}=67.77$ km s$^{-1}$ Mpc$^{-1}$. \label{Figure 2}}
\end{figure*}

\begin{figure*}
\caption{\textbf{Top:} Distribution of morphological types in \texttt{CGS}. \textbf{Bottom:} Distribution of morphological types in our volume-limited sample.
Our sample preserves the distribution of morphological types in \texttt{CGS}.\label{Figure 3}}
\end{figure*}

\begin{figure*}
\caption{The T-type histograms for our volume-limited sample and \texttt{CGS} are shown. Our sample preserves the T-type distribution in \texttt{CGS}. The differences between the densities
of each T-type are always less than 5\% \label{Figure 4}.}
\end{figure*}

\begin{figure*}
\caption{This is the illustration of the methodology for estimating the errors on the model parameters. The galaxy profile (red 
dots), the galaxy profile$+$sky standard deviation (black dots), and the galaxy profile$-$sky standard deviation (blue dots) are 
shown with their best model fits for each galaxy. The differences between the radial profiles from the observed galaxy and its models are also shown below each panel.
\textbf{Left} panel shows the S\'{e}rsic bulge models (dashed lines) for the elliptical galaxy (\texttt{NGC1439}). 
The red one refers to the S\'{e}rsic bulge model with $R_{e}=57.02$, $n=4.98$, and $\mu_{e}=22.65$, the black one refers to the model with $R_{e}=82.98$, $n=5.56$, 
and $\mu_{e}=23.28$, and the blue one refers to the model with $R_{e}=41.34$, $n=4.48$, and $\mu_{e}=22.10$. 
\textbf{Right} panel shows the total galaxy models (dashed lines) of S\'{e}rsic bulge models (dotted lines) and exponential disc models (dot-dashed lines) for the lenticular galaxy 
(\texttt{ESO208-G021}). The red dashed line refers to the model with $R_{e}=14.02$, $n=2.37$, $\mu_{e}=19.12$, $R_{disk}=77.77$, and $\mu_{0}=22.26$, the black 
one refers to the model with $R_{e}=14.39$, $n=2.40$, $\mu_{e}=19.16$, $R_{disk}=116.23$, and $\mu_{0}=22.33$, the blue one refers to the model with $R_{e}=12.61$, $n=2.20$, 
$\mu_{e}=18.95$, $R_{disk}=39.35$, and $\mu_{0}=21.44$. \label{Figure 5}}
\end{figure*}

\begin{figure*}
\caption{The surface brightness profiles for 30 elliptical galaxies are shown with the best-fit galaxy models. The dashed line is the best S\'{e}rsic fit to the bulge. The differences between 
the radial profiles from the observed galaxy and its model are also shown below each panel.\label{Figure 6}}
\end{figure*}

\begin{figure*}
\caption{The surface brightness profiles for 38 S0 galaxies are shown with the best-fit galaxy model (black dashed line). The pink dotted line is the S\'{e}rsic fit to the bulge, the blue dot-dashed 
line is the exponential disc fit. The differences between the radial profiles from the observed galaxy and its model are also shown below each panel.\label{Figure 7}}
\end{figure*}

\begin{figure*}
\caption{The S\'{e}rsic index histogram (blue dashed line) and the \texttt{PDF} fits (red solid line from \texttt{PEARSPDF}, pink solid line from 
\texttt{ALLFITDIST}) to the data are shown. The \texttt{PDF} (red solid line) from \texttt{PEARSPDF} is defined by the statistical standardized moments: $\mu=3.10$, $stdev=1.38$, 
$skewness=0.95$, and $kurtosis=4.17$. The \texttt{PDF} (pink solid line) from \texttt{ALLFITDIST} is a gamma distribution function with $\mu=3.11$, $variance=1.84$, shape $a=5.26$, 
scale $b=0.59$. \label{Figure 8}}
\end{figure*}

\begin{figure*}
\caption{\textbf{Left:} Impact of the uncertainties on the shape of the \texttt{BHMF} is shown, first assuming no errors, then allowing the listed errors to be perturbed individually and 
then collectively. The uncertainities in the S\'{e}rsic index distribution have no impact on the \texttt{BHMF} for $M_{bh}>10^{9}M_{\sun}$. The uncertainties in the $M_{bh}$-$n$ relation dominate 
at this region, softening the high-mass decrease of the \texttt{BHMF}, and thus increasing the total density of the \texttt{BHMF} for high masses. 
\textbf{Right:} Best estimate of the early-type \texttt{BHMF} is obtained by merging all the \texttt{MCMC} realizations of the \texttt{BHMF}s after allowing the listed errors to be 
perturbed collectively. The solid red line represents the $50th$ percentile and the green shaded region is delimited by the $16th$ and $84th$ percentile levels. Note that the \texttt{BHMF} 
estimates are all normalized by adding $-0.25$ in the y-axis to be able correct for the overdensity in our survey volume.  \label{Figure 9}}
\end{figure*}

\begin{figure*}
\caption{Best estimate of the \texttt{BHMF} is obtained by merging all the \texttt{MCMC} realizations of the \texttt{BHMF}s from the early- and late-type galaxies. The $M_{bh}$-$n$ and 
$M_{bh}$-$P$ relations are used for the early- and late-type galaxies, respectively. The \texttt{MCMC} sampling is used to account for the uncertainies from both
the measurements and the scaling relations. The all-type \texttt{BHMF} (red solid line) is defined by the $50th$ percentile, while its error region (green shaded region) is 
delimited by the $16th$ and $84th$ percentile levels of the merged \texttt{MCMC} realizations. The blue and pink dotted lines show the 1$\sigma$ uncertainity region for 
the late- and early-type \texttt{BHMF}, respectively. This clearly shows that the late-type \texttt{BHMF} dominates at the low-mass end while the early-type \texttt{BHMF} dominates at 
the high-mass end. Note that the \texttt{BHMF} estimates are all normalized by adding $-0.25$ in the y-axis.\label{Figure 10}}
\end{figure*}

\begin{figure*}
\caption{Comparison of our early-type \texttt{BHMF} (solid red line) with a green shaded $\pm1\sigma$ error region; with those of \citet{Graham2007} (pink triangles: GR07); \citet{Marconi2004} (black
stars: M04); and \citet{Vika2009} (blue open circles: V09). The \texttt{BHMF} data of \citet{Vika2009} below $\log(M_{bh}/M_{\sun}) = 7.67$, which \citet{Vika2009} considers unreliable
because it is derived from galaxies with $\mathfrak{M}_{B}>-18$, is depicted by the open circles with light blue color. Note that our \texttt{BHMF} is normalized by adding $-0.25$ 
in the y-axis to be able correct for the overdensity in our sample and all other \texttt{BHMF}s are transformed to $H_{0}=67.77$ km s$^{-1}$ Mpc$^{-1}$.\label{Figure 11}}
\end{figure*}

\begin{figure*}
\caption{Comparison of our determination of the \texttt{BHMF} (red solid line) for all galaxy types with a green shaded $\pm1\sigma$ error region;
with those of \citet{Graham2007} (pink triangles: GR07); \citet{Marconi2004} (black stars: M04) ; and \citet{Vika2009} (blue open circles: V09).
The \texttt{BHMF} data of \citet{Vika2009} below $\log(M_{bh}/M_{\sun}) = 7.67$, which \citet{Vika2009} considers unreliable
because it is derived from galaxies with $\mathfrak{M}_{B}>-18$, is depicted by the open circles with light blue color. Note that our \texttt{BHMF} is normalized by adding $-0.25$ 
in the y-axis to be able correct for the overdensity in our sample and all other \texttt{BHMF}s are transformed to $H_{0}=67.77$ km s$^{-1}$ Mpc$^{-1}$.\label{Figure 12}}
\end{figure*}

\begin{figure*}
\caption{Comparison of our determination of the \texttt{BHMF} (red solid line) for all galaxy types with a green shaded $\pm1\sigma$ error region with more recent works.
The blue solid lines show the 1$\sigma$ uncertainity region for the local \texttt{BHMF} from \citet{Shankar2013b}(S13), assuming the revised $M_{bh}$-$\sigma$ relation from \citet{McConnellMa2013} 
and applying it to all local galaxies. The region enclosed by the blue dashed lines is the same but assuming the \texttt{SMBH} mass in Sa galaxies is negligible. The black dot-dashed line 
shows the local \texttt{BHMF} derived by using the continuity equation models of \citet{Shankar2013a} and assuming a characteristic Eddington ratio decreasing with cosmological time. 
The pink dotted line marks the local \texttt{BHMF} in the Illustris simulated volume \citep{Sijacki2015}. Note that our \texttt{BHMF} is normalized by adding $-0.25$ 
in the y-axis to be able correct for the overdensity in our sample and all other \texttt{BHMF}s are transformed to $H_{0}=67.77$ km s$^{-1}$ Mpc$^{-1}$. \label{Figure 13}}
\end{figure*}

\clearpage
\begin{deluxetable}{crr} 
\tablecolumns{3} 
\tablewidth{0pc}
\tabletypesize{\scriptsize}
\tablecaption{BHMF VALUES} 
\tablehead{ 
\colhead{$\log M_{bh}/M_{\sun}$}    &  \multicolumn{2}{c}{$\log \varphi$ [$h^{3}_{67.77}$ $Mpc^{-3}$ $dex^{-1}$]} \\ 
\colhead{(1)} & \colhead{Early type (2)}   & \colhead{All galaxies (3)}}
\startdata 
$5.00$ &   $-4.42^{+0.27}_{-0.31}$ & $-3.84^{+0.30}_{-0.39}$ \\
$5.25$ &   $-4.29^{+0.25}_{-0.31}$ & $-3.66^{+0.24}_{-0.35}$ \\
$5.50$ &   $-4.18^{+0.24}_{-0.28}$ & $-3.48^{+0.22}_{-0.31}$ \\
$5.75$ &   $-4.08^{+0.22}_{-0.26}$ & $-3.30^{+0.19}_{-0.25}$ \\
$6.00$ &   $-3.96^{+0.20}_{-0.24}$ & $-3.14^{+0.14}_{-0.22}$ \\
$6.25$ &   $-3.84^{+0.17}_{-0.22}$ & $-3.00^{+0.11}_{-0.18}$\\
$6.50$ &   $-3.73^{+0.15}_{-0.19}$ & $-2.88^{+0.07}_{-0.13}$\\
$6.75$ &   $-3.62^{+0.13}_{-0.17}$ & $-2.80^{+0.06}_{-0.08}$\\
$7.00$ &   $-3.51^{+0.11}_{-0.14}$ & $-2.74^{+0.06}_{-0.07}$\\
$7.25$ &  $-3.41^{+0.08}_{-0.12}$ & $-2.72^{+0.08}_{-0.11}$\\
$7.50$ &  $-3.32^{+0.07}_{-0.09}$ & $-2.75^{+0.12}_{-0.15}$ \\
$7.75$ &   $-3.25^{+0.06}_{-0.06}$ & $-2.85^{+0.16}_{-0.17}$\\
$8.00$ &   $-3.21^{+0.07}_{-0.06}$ & $-2.97^{+0.17}_{-0.16}$\\
$8.25$ &   $-3.20^{+0.08}_{-0.09}$ & $-3.09^{+0.16}_{-0.15}$\\
$8.50$ &   $-3.27^{+0.11}_{-0.16}$ & $-3.25^{+0.15}_{-0.18}$\\
$8.75$ &   $-3.45^{+0.17}_{-0.27}$ & $-3.44^{+0.17}_{-0.28}$\\
$9.00$ &   $-3.71^{+0.25}_{-0.39}$ & $-3.70^{+0.25}_{-0.39}$\\ 
$9.25$ &   $-4.02^{+0.35}_{-0.46}$ & $-3.99^{+0.34}_{-0.50}$ \\
$9.50$ &   $-4.32^{+0.39}_{-0.64}$ & $-4.32^{+0.42}_{-0.59}$ \\
\enddata 
\tablecomments{Columns: (1) \texttt{SMBH} mass listed as $\log(M/M_{\sun})$ in 0.25 dex intervals. (2) Normalized \texttt{BHMF} data for early-type galaxies in our sample, as presented in
Figure 9 (left), in units of $h^{3}_{67.77} Mpc^{-3} dex^{-1}$. (3) Normalized \texttt{BHMF} data for all galaxies in our sample, as presented in Figure 10, in units of $h^{3}_{67.77} Mpc^{-3} dex^{-1}$. }
\end{deluxetable}

\clearpage
\begin{deluxetable}{lcccc}
\tablecolumns{5}
\tablewidth{0pt}
\tabletypesize{\scriptsize}
\tablecaption{Black Hole Mass Function Evaluation}
\tablehead{
 \colhead{N} &
 \colhead{$M_{Total}$} &
 \colhead{$\rho$} & 
 \colhead{$\Omega_{BH}$} &
 \colhead{$\Omega_{BH}/\omega_{b}$}\\
 \colhead{} &
 \colhead{($10^{10}$ $M_{\sun}$)} &
 \colhead{($10^{5}$ $h^{3}_{67.77}$ $M_{\sun}$ $Mpc^{-3}$)} &
 \colhead{($10^{-6}$ $h_{67.77}$)} &
 \colhead{($h^{3}_{67.77}$ $\%$)}\\
 \colhead{(1)} &
 \colhead{(2)} &
 \colhead{(3)} &
 \colhead{(4)} &
 \colhead{(5)}}
\startdata
68 (E/S0)\tablenotemark{b} & $1.05^{+0.47}_{-0.36}$ & $3.10^{+1.40}_{-1.06}$ & $2.44^{+1.10}_{-0.83}$ & $0.011^{+0.005}_{-0.004}$\\
68 (E/S0)\tablenotemark{a} & $0.59^{+0.26}_{-0.20}$ & $1.74^{+0.79}_{-0.60}$ & $1.37^{+0.62}_{-0.47}$ & $0.006^{+0.003}_{-0.002}$\\
140 (Spiral)\tablenotemark{*,b} & $0.18^{+0.22}_{-0.09}$ & $0.55^{+0.65}_{-0.27}$ & $0.43^{+0.51}_{-0.21}$ & $0.002^{+0.003}_{-0.002}$\\
140 (Spiral)\tablenotemark{a} & $0.10^{+0.12}_{-0.05}$ & $0.30^{+0.37}_{-0.15}$ & $0.24^{+0.29}_{-0.12}$ & $0.001^{+0.002}_{-0.001}$\\
208 (All-type)\tablenotemark{b} & $1.23^{+0.69}_{-0.45}$ & $3.65^{+2.05}_{-1.33}$ & $2.87^{+1.60}_{-1.11}$ & $0.013^{+0.008}_{-0.006}$\\
208 (All-type)\tablenotemark{a} & $0.69^{+0.38}_{-0.25}$ & $2.04^{+1.16}_{-0.75}$ & $1.61^{+0.91}_{-0.59}$ & $0.007^{+0.005}_{-0.003}$\\
\enddata
\tablenotetext{b}{Before the normalization is applied}
\tablenotetext{a}{After the normalization is applied}
\tablenotetext{*}{Data taken from \citet{Davis2014}}
\tablecomments{Columns: (1) Number of galaxies. (2) Total mass from the summation of all the \texttt{SMBH}s in units of $10^{10}$ $M_{\sun}$.
(3) Density of \texttt{SMBH}s in units of $10^{5}$ h$^{3}_{67.77}$ $M_{\sun}$ Mpc$^{-3}$. (4) Cosmological \texttt{SMBH} mass density [$\Omega_{BH}=\rho/\rho_{0}$,
assuming $\rho_0 = 3H^{2}_{0}/8 \pi \text{G}=1.274 \times 10^{11}$ $M_{\sun}$ Mpc$^{-3}$ when $H_0=67.77$ km s$^{-1}$ Mpc$^{-3}$]. 
(5) Fraction of the universal baryonic inventory locked up in \texttt{SMBH}s [$\Omega_{BH}/\omega_{b}$].}
\end{deluxetable}

\clearpage
\LongTables
\begin{deluxetable*}{llrcccccccr}
\tablecolumns{11}
\tabletypesize{\tiny}
\tablecaption{VOLUME LIMITED SAMPLE}
\tablehead{
   \colhead{Galaxy Name} &
   \colhead{Hubble Type} &
   \colhead{$D_{L}$} &
   \colhead{$B_{T}$} &
   \colhead{Band} &
   \colhead{$R_{e}/$\texttt{FWHM}}&
   \colhead{$n$} &
   \colhead{$\log(M/M_{\sun})$} &
   \colhead{Telescope}\\
   \colhead{(1)} &
   \colhead{(2)} &
   \colhead{(3)} &
   \colhead{(4)} &
   \colhead{(5)} &
   \colhead{(6)} &
   \colhead{(7)} &
   \colhead{(8)} &
   \colhead{(9)}
}
\startdata
ESO 208$-$G021 &S0\tablenotemark{*} &17.0 &$-19.57$  &R &14 &$2.37^{+0.03}_{-0.17}$  &$7.57^{+0.02}_{-0.14}$ &LCO\\
ESO 221$-$G026 &E &12.4 &$-19.13$ &H &30 &$4.11^{+0.58}_{-0.78}$  &$8.43^{+0.15}_{-0.29}$ &2MASS\\
ESO 311$-$G012 &S0/a &18.3  &$-20.24$  &H &6 &$1.91^{+0.55}_{-0.36}$  &$7.13^{+0.51}_{-0.48}$ &2MASS\\
IC 5181 &S0 &24.8 &$-19.38$ &H &2 &$1.35^{+0.19}_{-0.04}$ &$6.32^{+0.32}_{-0.08}$ &2MASS\\
NGC 584 &E &19.5 &$-20.07$ &H &29 &$4.08^{+0.83}_{-0.43}$ &$8.42^{+0.21}_{-0.15}$ &2MASS\\
NGC 596 &S0\tablenotemark{*} &20.6 &$-19.68$ &H &3 &$2.84^{+0.26}_{-0.71}$ &$7.89^{+0.14}_{-0.53}$ &2MASS\\
NGC 636 &E &25.4 &$-19.62$ &H &13 &$7.15^{+0.19}_{-0.91}$ &$8.93^{+0.02}_{-0.09}$ &2MASS\\
NGC 720 &E &23.9 &$-20.54$ &H &25 &$2.53^{+0.36}_{-0.28}$  &$7.69^{+0.23}_{-0.22}$ &2MASS\\
NGC 936 &S0/a &20.7 &$-20.55$ &H &10 &$2.63^{+0.22}_{-0.45}$ &$7.76^{+0.13}_{-0.35}$ &2MASS\\
NGC 1052 &E &19.6 &$-19.93$ &3.6\micron &14 &$2.99^{+0.49}_{-0.26}$ &$7.97^{+0.23}_{-0.15}$ &Spitzer\\
NGC 1172 &E &24.3 &$-19.30$ &F160W &41 &$3.63^{+0.19}_{-0.18}$ &$8.26^{+0.07}_{-0.07}$ &HST\\
NGC 1201 &S0\tablenotemark{*} &20.2 &$-19.52$ &R &5 &$1.38^{+0.10}_{-0.06}$ &$6.38^{+0.34}_{-0.31}$ &LCO\\
NGC 1291 &S0/a &8.6 &$-20.12$ &H &9 &$1.60^{+0.20}_{-0.13}$ &$6.74^{+0.27}_{-0.20}$ &2MASS\\
NGC 1302 &S0/a &20.0 &$-19.62$ &3.6\micron &11 &$3.29^{+0.11}_{-0.17}$ &$8.12^{+0.05}_{-0.08}$ &Spitzer\\
NGC 1316 &S0 &19.2 &$-21.71$ &H &23 &$2.32^{+0.94}_{-0.15}$ &$7.65^{+0.53}_{-0.12}$ &2MASS\\
NGC 1326 &S0/a &16.9 &$-19.68$ &3.6\micron &7 &$2.21^{+0.21}_{-0.05}$ &$7.47^{+0.17}_{-0.04}$ &Spitzer\\
NGC 1332 &E\tablenotemark{*} &18.9 &$-20.11$ &3.6\micron &20 &$4.05^{+0.25}_{-0.25}$ &$8.41^{+0.23}_{-0.23}$ &Spitzer\\
NGC 1340 &E &18.2 &$-20.14$ &H &42 &$4.04^{+0.92}_{-0.62}$  &$8.41^{+0.23}_{-0.23}$ &2MASS\\
NGC 1351 &E\tablenotemark{*} &20.4 &$-19.17$ &F160W &32 &$2.57^{+0.14}_{-0.13}$  &$7.72^{+0.09}_{-0.10}$ &HST\\
NGC 1374 &E &19.4 &$-19.46$ &H &12 &$4.59^{+0.18}_{-0.84}$  &$8.56^{+0.04}_{-0.25}$ &2MASS\\
NGC 1379 &E &18.1 &$-19.39$ &H &14 &$3.56^{+0.21}_{-0.27}$  &$8.24^{+0.08}_{-0.11}$ &2MASS\\
NGC 1380 &S0 &18.0 &$-20.29$ &K &12 &$2.02^{+0.53}_{-0.12}$  &$7.26^{+0.45}_{-0.13}$ &2MASS\\
NGC 1387 &S0\tablenotemark{*} &18.0 &$-19.56$ &H &3 &$2.72^{+0.56}_{-0.46}$ &$7.82^{+0.30}_{-0.34}$ &2MASS \\
NGC 1395 &E &21.9 &$-21.11$ &H &33 &$3.68^{+0.39}_{-0.63}$  &$8.28^{+0.13}_{-0.28}$ &2MASS\\
NGC 1399 &E &18.9 &$-21.10$ &H &34 &$4.63^{+0.63}_{-1.07}$  &$8.57^{+0.13}_{-0.33}$ &2MASS\\
NGC 1400 &E\tablenotemark{*} &23.6 &$-19.81$ &F160W &21 &$2.09^{+0.05}_{-0.07}$  &$7.32^{+0.05}_{-0.07}$ &HST\\
NGC 1404 &S0\tablenotemark{*} &18.6 &$-20.49$ &3.6\micron &10 &$2.86^{+0.11}_{-0.62}$  &$7.90^{+0.06}_{-0.44}$ &Spitzer\\
NGC 1407 &E &23.8 &$-21.22$ &H &45 &$4.39^{+0.91}_{-0.80}$ &$8.51^{+0.20}_{-0.26}$ &2MASS\\
NGC 1427 &E &20.9 &$-19.80$ &R &52 &$5.02^{+0.83}_{-0.76}$  &$8.65^{+0.14}_{-0.18}$ &LCO\\
NGC 1439 &E &24.3 &$-19.65$ &R &56 &$4.98^{+0.58}_{-0.50}$ &$8.65^{+0.10}_{-0.11}$ &LCO\\
NGC 1452 &S0/a &22.8 &$-19.24$ &4.5\micron &5 &$1.62^{+0.11}_{-0.04}$ &$6.77^{+0.15}_{-0.06}$ &Spitzer\\
NGC 1527 &S0\tablenotemark{*} &16.7 &$-19.49$ &J &6 &$1.91^{+0.23}_{-0.22}$ &$7.14^{+0.24}_{-0.28}$ &2MASS\\
NGC 1533 &S0\tablenotemark{*} &18.4 &$-19.56$ &3.6\micron &3 &$1.23^{+0.04}_{-0.03}$  &$6.09^{+0.07}_{-0.07}$ &Spitzer\\
NGC 1537 &E\tablenotemark{*} &18.8 &$-19.82$ &H &16 &$2.89^{+0.86}_{-0.37}$ &$7.92^{+0.39}_{-0.24}$ &2MASS\\
NGC 1543 &S0 &17.5 &$-19.78$ &H &7 &$1.15^{+0.04}_{-0.26}$  &$5.91^{+0.09}_{-0.72}$ &2MASS\\
NGC 1549 &E &16.4 &$-20.43$ &H &23 &$5.62^{+1.21}_{-0.91}$  &$8.76^{+0.15}_{-0.17}$ &2MASS\\
NGC 1553 &S0 &14.6 &$-20.57$ &4.5\micron &9 &$1.78^{+0.32}_{-0.81}$  &$7.73^{+0.20}_{-0.75}$ &Spitzer\\
NGC 1574 &S0\tablenotemark{*} &18.6 &$-20.05$ &H &8 &$1.68^{+0.06}_{-0.15}$  &$6.85^{+0.08}_{-0.22}$ &2MASS\\
NGC 2217 &S0/a &19.5 &$-20.04$ &H &5 &$1.24^{+0.04}_{-0.46}$  &$6.11^{+0.08}_{-1.33}$ &2MASS\\
NGC 2325 &E &22.6 &$-19.82$ &R &54 &$2.73^{+0.35}_{-0.15}$  &$7.82^{+0.20}_{-0.10}$ &LCO\\
NGC 2380 &S0 &22.2 &$-20.54$ &H &10 &$3.28^{+0.10}_{-0.05}$  &$8.12^{+0.04}_{-0.02}$ &2MASS\\
NGC 2434 &E &21.9 &$-20.23$ &H &14 &$5.03^{+0.97}_{-1.34}$  &$8.65^{+0.16}_{-0.37}$ &2MASS\\
NGC 2640 &E\tablenotemark{*} &17.2 &$-20.13$ &H &26 &$3.82^{+0.47}_{-0.40}$ &$8.33^{+0.15}_{-0.15}$ &2MASS\\
NGC 2784 &S0 &8.5 &$-19.22$ &H &15 &$3.79^{+0.54}_{-0.69}$  &$8.32^{+0.17}_{-0.29}$ &2MASS\\
NGC 2822 &S0\tablenotemark{*} &24.7 &$-20.62$ &R &6 &$1.60^{+0.01}_{-0.03}$  &$6.74^{+0.01}_{-0.04}$ &LCO\\
NGC 2974 &S0\tablenotemark{*} &25.3 &$-20.30$ &R &18 &$3.23^{+0.40}_{-0.22}$ &$8.10^{+0.17}_{-0.11}$ &LCO\\
NGC 3115 &S0\tablenotemark{*} &10.1 &$-20.09$ &3.6\micron &11 &$2.11^{+0.58}_{-0.39}$ &$7.34^{+0.46}_{-0.44}$ &Spitzer\\
NGC 3136 &E &23.9 &$-20.93$ &R &92 &$4.77^{+0.74}_{-0.56}$  &$8.60^{+0.14}_{-0.14}$ &LCO\\
NGC 3585 &E &17.6 &$-20.63$ &H &35 &$3.85^{+0.91}_{-0.68}$ &$8.34^{+0.25}_{-0.28}$ &2MASS\\
NGC 3904 &E &24.7 &$-20.28$ &3.6\micron &20 &$3.23^{+0.77}_{-0.23}$ &$8.10^{+0.30}_{-0.12}$ &Spitzer\\
NGC 3923 &E &20.9 &$-21.13$ &H &19 &$3.12^{+0.60}_{-0.09}$ &$8.04^{+0.26}_{-0.05}$ &2MASS\\
NGC 3955 &S0/a &20.6 &$-19.15$ &3.6\micron &12 &$2.50^{+0.10}_{-0.53}$ &$7.67^{+0.07}_{-0.46}$ &Spitzer\\
NGC 4024 &E\tablenotemark{*} &25.4 &$-19.34$ &H &12 &$4.74^{+5.20}_{-0.80}$ &$8.59^{+0.47}_{-0.22}$ &2MASS\\
NGC 4546 &S0\tablenotemark{*} &17.3 &$-19.74$ &3.6\micron &8 &$2.51^{+0.20}_{-0.20}$ &$7.67^{+0.14}_{-0.16}$ &Spitzer\\
NGC 4684 &S0/a &20.5 &$-19.38$ &H &7 &$3.44^{+0.20}_{-0.20}$ &$8.19^{+0.08}_{-0.09}$ &2MASS \\
NGC 4691 &S0/a &22.5 &$-19.83$ &3.6\micron &10 &$1.12^{+0.02}_{-0.09}$  &$5.82^{+0.05}_{-0.23}$ &Spitzer\\
NGC 4697 &E &11.6 &$-20.00$ &H &47 &$3.02^{+0.38}_{-0.32}$  &$7.99^{+0.18}_{-0.19}$ &2MASS\\
NGC 4753 &S0 &19.6 &$-20.46$ &3.6\micron &16 &$3.86^{+0.38}_{-0.15}$ &$8.35^{+0.12}_{-0.05}$ &Spitzer\\
NGC 4856 &S0/a &21.1 &$-20.36$ &K &13 &$2.79^{+0.33}_{-0.29}$  &$7.86^{+0.18}_{-0.19}$ &2MASS\\
NGC 4958 &S0 &20.9 &$-20.03$ &3.6\micron &10 &$2.25^{+0.13}_{-0.04}$ &$7.47^{+0.11}_{-0.03}$ &Spitzer\\
NGC 4976 &E &12.5 &$-20.20$ &J &58 &$4.22^{+0.61}_{-0.58}$  &$8.46^{+0.15}_{-0.19}$ &2MASS\\
NGC 4984 &S0\tablenotemark{*} &21.3 &$-19.62$ &4.5\micron &9 &$4.46^{+0.09}_{-1.27}$ &$8.53^{+0.02}_{-0.45}$ &Spitzer\\
NGC 5128 &S0 &3.7 &$-20.53$ &H &20 &$1.54^{+0.44}_{-0.37}$ &$6.65^{+0.56}_{-0.70}$ &2MASS\\
NGC 6684 &S0 &12.4 &$-19.4$ &H &4 &$2.60^{+0.25}_{-0.39}$  &$7.74^{+0.16}_{-0.30}$ &2MASS\\
NGC 7041 &S0\tablenotemark{*} &24.9 &$-19.81$ &R &7 &$1.85^{+0.34}_{-0.23}$ &$7.07^{+0.35}_{-0.30}$  &LCO\\
NGC 7144 &S0\tablenotemark{*} &24.9 &$-20.35$ &H &3  &$2.42^{+0.34}_{-0.23}$ &$7.61^{+0.23}_{-0.19}$ &2MASS\\
NGC 7145 &S0\tablenotemark{*} &23.1 &$-19.32$ &R &11 &$2.45^{+0.10}_{-0.35}$ &$7.63^{+0.07}_{-0.30}$ &LCO\\
NGC 7507 &E &22.5 &$-20.35$ &K &16 &$5.83^{+1.33}_{-0.64}$ &$8.79^{+0.15}_{-0.10}$ &2MASS\\
\enddata
\tablenotetext{*}{Hubble type for this galaxy is determined based on its light profile.}
\tablecomments{Columns: (1) Galaxy name. (2) Hubble type, from http://cgs.obs.carnegiescience.edu/CGS/database\_tables.
(3) Luminosity distance in Mpc, compiled from the mean redshift-independent distance from \texttt{NED}.
(4) B-band absolute magnitude, determined from formula: $\mathfrak{M}_{B}=B_{T}-5\log(D_{L})+5-A_{B}$, where $B_{T}$ is total B-band apparent magnitude (taken from \textit{HyperLeda}), 
$A_{B}$ is galactic extinction in B-band (from \citet{SchlaflyFinkbeiner2011}, as compiled by the \texttt{NED}), and $D_{L}$ is luminosity distance in units of $pc$. (5) Band.
(6) $R_{e}/$\texttt{FWHM} ratio. (7) S\'{e}rsic Index. (8) \texttt{SMBH} mass in $\log(M/M_{\sun})$,
converted from the S\'{e}rsic index via the Equation 1.  (9) Telescope from which the imaging data was taken.}
\end{deluxetable*}

\end{document}